\documentclass[prl,aps,twocolumn,floatfix,a4paper]{revtex4}

\usepackage{amsmath,amssymb,amsfonts}
\usepackage{bm}
\usepackage{graphicx}
\usepackage{mathrsfs}

\begin{document}

\title{Breakup of the Fermi surface near the Mott transition in low-dimensional
systems}

\author{C. Berthod and T. Giamarchi}
\affiliation{DPMC, Universit\'e de Gen\`eve, 24 quai Ernest-Ansermet,
1211 Gen\`eve 4, Switzerland}
\author{S. Biermann and A. Georges}
\affiliation{Centre de Physique Th\'eorique, CNRS-UMR 7644, Ecole Polytechnique,
91128 Palaiseau Cedex, France}

\begin{abstract}

We investigate the Mott transition in weakly-coupled one-dimensional (1d)
fermionic chains. Using a generalization of Dynamic Mean Field Theory, we show
that the Mott gap is suppressed at some critical hopping $t_{\perp}^{c2}$. The
transition from the 1d insulator to a 2d metal proceeds through an intermediate
phase where the Fermi surface is broken into electron and hole pockets. The
quasiparticle spectral weight is strongly anisotropic along the Fermi surface,
both in the intermediate and metallic phases. We argue that such pockets would
look like `arcs' in photoemission experiments.

\end{abstract}

\date{\today}
\maketitle

The formation of a pseudogap phase near the Mott transition in
strongly correlated electron system is a long-standing question in condensed
matter physics. Since the observation of a pseudogap state in high-$T_c$
copper-oxide superconductors by a variety of experimental techniques
\cite{timusk_review_pseudogap}, the nature and origin of this phase lies at the
core of a heated debate. Several related issues, such as the possible breakup of
the Fermi surface (FS) into arcs or pockets \cite{norman_fermi_arcs,
damascelli_review_ARPES} and the concomitant appearance of `hot spots' in the
quasiparticle spectrum, have also attracted much attention in recent years.
The theoretical description of the Mott transition in two spatial dimensions has
proven to be difficult \cite{imada_mott_review}, however, and a consensus
on the nature and properties of the pseudogap has not yet emerged. The central
question concerns the way the FS is destroyed when entering the Mott state,
either at zero temperature as a function of some control parameter, or with
decreasing temperature from a high-$T$ metallic phase.

For solving these questions one
has to resort to approximate or numerical approaches. Among those methods, the
Dynamic Mean Field Theory (DMFT) \cite{georges_d=infini,kotliar_dmft_review} has
proven to be very fruitful to tackle the question of Mott transitions and strong
correlations in high dimensional systems. However, since in the standard
implementation of this method one reduces the problem to a single site, the
momentum dependence of the self energy is lost, which makes it unable to
describe the FS anisotropies, and answer the above questions. To overcome this
limitation, various cluster-DMFT schemes have been used to approach the 2d
problem \cite{maier_review_RMP, tremblay_review_pseudogap, kotliar_review,
parcollet_hotspots_dmft}. One difficulty with these approaches which, for small
clusters, can affect physical predictions, lies in a certain degree of
arbitrariness in converting the cluster results into physical quantities on the
lattice.

In addition to high-$T_c$ superconductors, these questions related to the
approach of the Mott transition are directly relevant to other systems in which
interactions are deemed to be important such as the organic superconductors,
made of weakly-coupled one-dimensional chains \cite{giamarchi_review_chemrev}.
Such systems exhibit a deconfinement transition between a one-dimensional Mott
insulator and a three dimensional metal. For such systems of coupled chains, FS
pockets were found when the inter-chain coupling is treated within the RPA
approximation \cite{essler_rpa_quasi1d}. In that model the pockets form when
the inter-chain bandwidth exceeds the 1d Mott gap of the chains. However the RPA
approximation neglects the feedback of the inter-chain hopping on the 1d Mott
gap, and more generally on the self-energy. Depending on the evolution of the
Mott gap with inter-chain coupling, the FS pockets may or may not form. To go
beyond this approximation, we have proposed a generalization of the one-site
DMFT to one-dimensional chains embedded in a self-consistent bath
\cite{georges_organics_dinfiplusone}. This ch-DMFT scheme was successfully
applied to study the deconfinement transition in quasi-1d fermionic lattices,
and is the ideal tool to investigate the Mott transition in this type of systems
\cite{biermann_dmft1d_hubbard_short, biermann_oned_crossover_review,
giamarchi_iscom_1d}. This method has the advantage of having no arbitrariness in
the definition of the cluster, since here the cluster corresponds to a
physically well-defined entity (a single chain).

In the present paper, we use this method to study a two-dimensional lattice of
spinless fermions made of weakly coupled one-dimensional chains. Contrarily to
the 1d Hubbard model, the 1d spinless model has a Mott transition for a finite
value of the interaction. Furthermore, the effects of the interaction are
stronger in the spinless case than in the spin-$\frac{1}{2}$ case, as
illustrated by the lower value of the Luttinger coefficient which can be reached
in the former. This makes the spinless model very well suited to tackle these
issues. The spinless model can also be viewed as a caricature of a
spin-$\frac{1}{2}$ model at quarter filling with a very strong local repulsion.
Using the ch-DMFT method we show that FS pockets exist in this model close to
the metal-insulator transition. However, due to the very strong anisotropy of
the spectral function and of the spectral weight along the Fermi surface, these
pockets would appear as Fermi-surface arcs in experiments. We note that Stanescu
and Kotliar \cite{stanescu_pockets_cluster_dmft} and Kuchinskii \textit{et. al.}
\cite{kuchinskii_2005} have discussed similar effects using extensions of DMFT
including momentum dependence.

Our model has two parameters, $t_{\perp}$ and $V$, which control the inter-chain
hopping and the interaction $\mathcal{H}_{\text{int}}=V\sum_r n_rn_{r+1}$ within
the chains, respectively, relative to the in-chain hopping $t$. In the 1d limit
($t_{\perp}=0$) this model has a phase transition between a Luttinger liquid at
$V<2\,t$ and a Mott insulator at $V>2t$. Here we focus on the region $V>2\,t$,
and we investigate the destruction of the Mott insulator with increasing
$t_{\perp}$. Within ch-DMFT the 2d problem is mapped onto an effective 1d
problem described by the action
	\begin{eqnarray}\label{eq:action}
		\nonumber
		\mathcal{S}_{\text{eff}}&=&-\sum_{rr'}\int_0^{\beta}d\tau d\tau'\,
			c^{\dagger}_r(\tau)\mathscr{G}_0^{-1}(r-r',\tau-\tau')
			c^{\phantom{\dagger}}_{r'}(\tau')\\
			&&+\int_0^{\beta}d\tau\,\mathcal{H_{\text{int}}}.
	\end{eqnarray}
The inverse propagator $\mathscr{G}_0^{-1}$ in Eq.~(\ref{eq:action}) plays the
role of a long-range, time-dependent hopping amplitude. It must be determined
from the requirement that the in-chain Green's function $\mathscr{G}$ calculated
from $\mathcal{S}_{\text{eff}}$ coincides with the $k_{\perp}$-summed Green's
function of the original 2d model \cite{georges_d=infini,
biermann_dmft1d_hubbard_short}. This requirement implies
	\begin{equation}\label{eq:DMFT}
		\mathscr{G}_0(k,\omega)=\frac{1}{\omega-\xi_k+
		\mathscr{G}^{-1}(k,\omega)-R[\mathscr{G}(k,\omega)]},
	\end{equation}
with $k$ the in-chain momentum, $\xi_k=-2t\cos k-\mu$ the bare dispersion, and
$R(z)=\text{sign}[\text{Re}(z)]\sqrt{1/z^2+(2t_{\perp})^2}$ resulting from the
integration over the transverse energy
$\varepsilon_{\perp}=-2t_{\perp}\cos(k_{\perp})$. Eqs.~(\ref{eq:action}) and
(\ref{eq:DMFT}) can be readily derived from the assumption that the lattice
self-energy does not depend on transverse momentum $k_{\perp}$, and is given by
the effective in-chain self-energy:
	\begin{equation}\label{eq:Sigma}
		\Sigma(k,\omega)
		=\mathscr{G}_0^{-1}(k,\omega)-\mathscr{G}^{-1}(k,\omega).
	\end{equation}
$\Sigma(k,\omega)$ will be our main concern here. In order to evaluate the
self-energy we compute the space-time Green's function
$\mathscr{G}(r,\tau)=-\langle c^{\phantom{\dagger}}_r(\tau)
c^{\dagger}_0(0)\rangle$ of the action Eq.~(\ref{eq:action}) by quantum
Monte-Carlo on a discrete imaginary-time mesh $\tau_{\ell}=\ell\beta/L$, using
the Hirsch-Fye algorithm \cite{hirsch_fye_qmc}. From the Fourier transform
$\mathscr{G}(k,i\omega_n)$ we construct a new propagator $\mathscr{G}_0$
according to Eq.~(\ref{eq:DMFT}), which is fed back into Eq.~(\ref{eq:action})
until self-consistency is achieved.

We consider a half-filled 32-sites chain closed with antiperiodic boundary
conditions. These boundary conditions were adopted for two reasons: (i) in short
1d chains we observed that the convergence of the Monte-Carlo toward
exact-diagonalization results is much faster with antiperiodic than with
periodic boundary conditions; (ii) for a given system size, the antiperiodic
boundary conditions improve the resolution near $k=\pi/2$, which is an advantage
for the investigation of the FS properties. Indeed, the shape of the FS is
controlled by the real part of the self-energy through the equation
	\begin{equation}\label{eq:FS_equation}
		\xi_k-2t_{\perp}\cos(k_{\perp})=-\text{Re}\,\Sigma(k,i0^+)
	\end{equation}
which must be solved at zero temperature near $k=\pi/2$. At finite temperature
the FS looses its identity, although sharp signatures may subsist in the
zero-energy spectral function $A(k,k_{\perp},\omega=0)$, which is accessible
through photoemission experiments. We will first discuss the FS topology and
properties implied by Eq.~(\ref{eq:FS_equation}) and our ch-DMFT results for
$\Sigma(k,\omega)$, before addressing some issues related to the experimental
measurement of the FS.

\begin{figure}[tb]
\includegraphics[width=7cm]{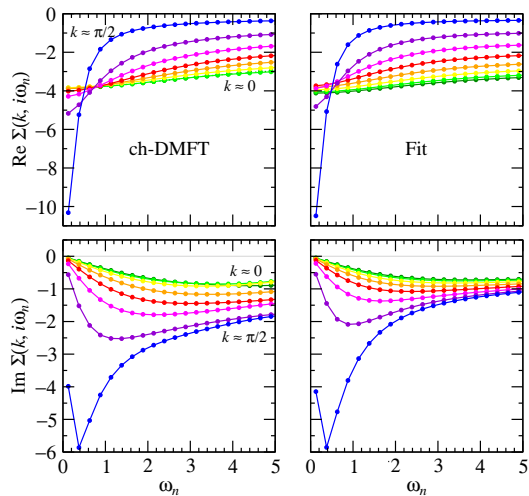}
\caption{\label{fig:fit}
Left panels: Self-energy on the imaginary-frequency axis for $k$ between $0$ and
$\pi/2$ at $V=4$, $t_{\perp}=0.5$, and $T=0.04$. The dots show the numerical
data and the lines are guides to the eye. The number of imaginary-time slices
was $L=60$. Right panels: Fit of the numerical data to a trial self-energy (see
text). All energies are in units of $t$.
}
\end{figure}

\begin{figure}[tb]
\includegraphics[width=8.6cm]{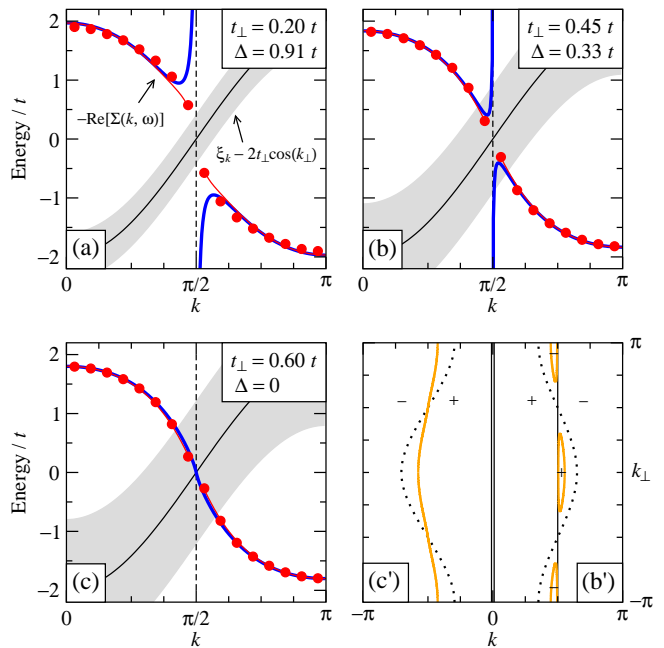}
\caption{\label{fig:pockets}
(a--c) Real part of the ch-DMFT self-energy as a function of $k$ at the lowest
Matsubara frequency for $V=2.5$, $T=0.1$, and increasing $t_{\perp}$ (red
points). The red and blue lines show the fit of the data to the function
$\Sigma_{\Delta}(k,i\omega_0)+\Sigma_{\text{int}}(k,i\omega_0)$, and its
continuation to zero energy, respectively. For the fit the model self-energy was
evaluated on a 32 $k$-point mesh, and then reevaluated using the fitted
parameters on a dense mesh at $\omega=i\omega_0$ and $T=0.1$ (red lines) and at
$\omega=i0^+$ and $T=0$ (blue lines). The shaded areas show the domain covered
by the free dispersion $\xi_k-2t_{\perp}\cos(k_{\perp})$ in
Eq.~(\ref{eq:FS_equation}). The Fermi surface corresponding to (b) and (c) is
shown in (b') and (c'), respectively. The dotted line indicates the
non-interacting Fermi surface in each case, and the $+$ and $-$ indicate the
sign of $\text{Re}\,\mathscr{G}(k,k_{\perp},0)$ within the Brillouin zone.
}
\end{figure}

For evaluating Eq.~(\ref{eq:FS_equation}) one needs a procedure to continue the
ch-DMFT self-energy from the lowest Matsubara frequency $i\omega_0=i\pi T$ down
to $\omega=i0^+$ along the imaginary-frequency axis. This is a delicate
endeavor, which in general requires some assumption about the analytical
behavior of $\Sigma(k,\omega)$ near $\omega=0$. We performed the continuation by
fitting the self-energy to an analytical function. The prominent feature in the
Mott phase is the spectral gap which can be crudely represented by a self-energy
of the form $\Sigma_{\Delta}(k,\omega) = (\lambda-1)\xi_k +
(\Delta/2)^2/(\omega+\lambda\xi_k)$, where $\Delta$ is the gap width and
$\lambda$ accounts for the dispersion renormalization due to exchange. A similar
Ansatz was recently proposed to describe high-$T_c$ cuprate superconductors
\cite{yang_2006}. This simple form is not sufficient to reproduce our QMC
results, however, even in the pure Mott phase at $t_{\perp}=0$. We obtained a
much better agreement with our data by taking into account residual
interactions. Specifically, the self-energy to which we fit the QMC results
takes the form $\Sigma(k,\omega) = \Sigma_{\Delta}(k,\omega) +
\Sigma_{\text{int}}(k,\omega)$, where $\Sigma_{\text{int}}$ contains all
diagrams---evaluated using the \emph{gapped} electron propagator $G_0(k,\omega)
= [\omega-\xi_k - \Sigma_{\Delta}(k,\omega)]^{-1}$ and an effective interaction
strength $V^*$---up to second order in perturbation theory \cite{lambda}. A
comparison of the QMC and model self-energies is displayed in
Fig.~\ref{fig:fit}. One can see that the model has enough freedom to fit the QMC
results in great detail, especially in the low-frequency region we are mostly
interested in. Furthermore, it turns out that the model fits the QMC data in the
whole range of temperature and $t_{\perp}$ values which we have investigated. We
can therefore use this fit to track the closing of the Mott gap and the
formation of the Fermi surface as $t_{\perp}$ is increased. A few additional
illustrations of the fit performance can be seen in Fig.~\ref{fig:pockets}.

In Fig.~\ref{fig:pockets} we display our results for
$-\text{Re}\,\Sigma(k,i\omega_0)$ at $V=2.5\,t$ and different values of
$t_{\perp}$, together with the fit results evaluated at $i\omega_0$ and $i0^+$.
The main trend with increasing $t_{\perp}$ can be seen on the raw numerical
data. For small $t_{\perp}$ the self-energy has a tendency to diverge near
$k_{\text{F}}=\pi/2$ (Fig.~\ref{fig:pockets}a). This behavior is most clearly
seen at large $V$ and low $T$ (see Fig.~\ref{fig:fit}). The singularity of
$\text{Re}\Sigma(k,0)$ results from the presence of a gap in the
zero-temperature spectral function at $k_{\text{F}}$ \cite{stanescu_preprint},
and is well captured by the model self-energy evaluated at $T=0$ (blue line). At
finite frequency and/or temperature the singularity is regularized as shown by
the red line. With increasing $t_{\perp}$ the drop of the QMC self-energy across
$k_{\text{F}}$ diminishes (Fig.~\ref{fig:pockets}b). Correspondingly the fitted
spectral gap $\Delta$ decreases and eventually vanishes at $t_{\perp}^{c2}\sim
0.5\,t$, together with the disappearance of the singularity in the self-energy
(Fig.~\ref{fig:pockets}c).

Fig.~\ref{fig:pockets} provides the graphical solution of
Eq.~(\ref{eq:FS_equation}), and illustrates the formation of the Fermi-surface
pockets in this model. When $t_{\perp}$ is small, the transverse dispersion is
not sufficient to overcome the gap in the self-energy, and the system remains
insulating. At high $t_{\perp}$, on the contrary, there is no gap in the
self-energy and Eq.~(\ref{eq:FS_equation}) has a solution for all $k_{\perp}$,
leading to a continuous Fermi surface. The latter has practically the same shape
as the non-interacting FS, but is strongly renormalized to an effective
inter-chain hopping $t_{\perp}^*\approx0.41\,t_{\perp}$. In the intermediate
regime $t_{\perp}^{c1}<t_{\perp}<t_{\perp}^{c2}$, there is a finite range of
$k_{\perp}$ values where Eq.~(\ref{eq:FS_equation}) admits two solutions,
leading to the breakup of the Fermi-surface into pockets
(Fig.~\ref{fig:pockets}b and b').

According to Luttinger's theorem, the area of the Brillouin zone where
$\text{Re}\,\mathscr{G}(k,k_{\perp},0)>0$ equals the electron density and is
thus conserved \cite{dzyaloshinskii_luttinger_surface, essler_rpa_quasi1d}. In
the Mott phase $\text{Re}\,\mathscr{G}(k,k_{\perp},0)$ changes sign at
$k=\pm\pi/2$ due to the divergence of $\text{Re}\,\Sigma(k,0)$
(Fig.~\ref{fig:pockets}a), and is positive in the domain $|k|<\pi/2$, leading to
a density $n=1/2$ as expected. Because the singularity of
$\text{Re}\,\Sigma(\pm\pi/2,0)$ subsists as long as $\Delta>0$, the line of
zeros of $\text{Re}\,\mathscr{G}(k,0)$ at $k=\pm\pi/2$ is still present when FS
pockets develop, as indicated in Fig.~\ref{fig:pockets}b'. On the other hand,
owing to particle-hole symmetry---which reduces at $\omega=0$ to an inversion
about the point $(k,k_{\perp})=(\pi/2,\pi/2)$---the electron and hole pockets
have identical volumes, so that Luttinger theorem is trivially obeyed in our
results.

The mechanisms of FS pockets formation in the present study and in the RPA
approach of Ref.~\cite{essler_rpa_quasi1d} are similar, although there is one
important difference. In the RPA approximation the spectral gap keeps its 1d
value at all $t_{\perp}$: pockets form when
$t_{\perp}>t_{\perp}^{c1}\sim\Delta_{\text{1d}}$, and they never merge into a
connected Fermi surface as $t_{\perp}$ continues increasing. Within ch-DMFT, in
contrast, the closing of the Mott gap with increasing $t_{\perp}$ is correctly
captured; as a result the pockets form at lower $t_{\perp}$ values---thus they
are very thin---and eventually they disappear at $t_{\perp}^{c2}$ where
$\Delta=0$.

\begin{figure}[tb]
\includegraphics[width=8.6cm]{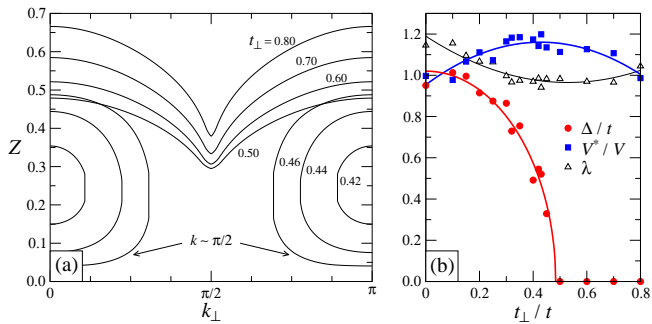}
\caption{\label{fig:hot_spots}
(a) Evolution of the quasiparticle residue $Z$ along the Fermi surface for
different inter-chain couplings $t_{\perp}$. When FS pockets are present there
are two values of $Z$ for each $k_{\perp}$, the lowest value corresponding to
the region of the pocket closest to $k=\pi/2$. (b) Parameters of the model
self-energy determined from fits to the ch-DMFT numerical data.
}
\end{figure}

The quasiparticle properties are strongly anisotropic along the Fermi-surface
pockets. It is already clear from Fig.~\ref{fig:pockets}b that the spectral
weight is much smaller on the vertical parts of the pockets closest to
$k=\pi/2$, due to the diverging self-energy in this region; as a result the
pockets would most likely look like `arcs' in photoemission experiments (see
below). Fig.~\ref{fig:hot_spots}a shows the evolution of the quasiparticle
residue along the Fermi surface. The residue was evaluated as
$Z=[1-d\text{Re}\,\Sigma(k,\omega)/d\omega|_{\omega=0}]^{-1}$ using the model
self-energy and an interpolation of the parameters fitted to the ch-DMFT data in
the whole range of $t_{\perp}$ values (Fig.~\ref{fig:hot_spots}b). In the
intermediate phase the residue on the vertical parts of the pockets decreases
from $Z\sim0.15$ to $Z\sim0$ with increasing $t_{\perp}$. On the `cold' side of
the pockets the behavior is inverted, and the residue increases from $Z\sim0.35$
to $Z\sim0.5$. Strikingly, a hot spot around $k=k_{\perp}=\pi/2$ subsists at
$t_{\perp}>t_{\perp}^{c2}$. Here again the evolution of $Z$ with $t_{\perp}$ is
different at the cold and hot spots: while $Z$ steadily approaches one in the
cold region, it remains close to $Z\sim0.4$ at the hot spots.

It's worth stressing the role of residual interactions in the results of
Figs.~\ref{fig:pockets} and \ref{fig:hot_spots}a. At the qualitative level, the
self-energy $\Sigma_{\Delta}(k,\omega)$ (with $\lambda=1$) is sufficient to
understand the formation of FS pockets with anisotropic residues. Indeed, using
$\Sigma=\Sigma_{\Delta}$ in Eq.~(\ref{eq:FS_equation}) one finds that pockets
form for $0<\Delta<2t_{\perp}$, and that the residue varies on such pockets.
However the pockets obtained in this way are considerably wider than in
Fig.~\ref{fig:pockets}, and the residue in the cold regions is $Z\sim1$ when
$\Delta$ approaches zero instead of $Z\sim0.5$ as in Fig.~\ref{fig:hot_spots}.
Thus the residual interactions are important for the quantitative understanding
of the FS properties. Meanwhile, the fact that our model self-energy fits the
ch-DMFT data at $t_{\perp}>t_{\perp}^{c2}$ with $V^*\sim V$
(Fig.~\ref{fig:hot_spots}b) indicates that second-order perturbation theory is a
good approximation in this region, as expected in a Fermi liquid.

\begin{figure}[tb]
\includegraphics[width=7cm]{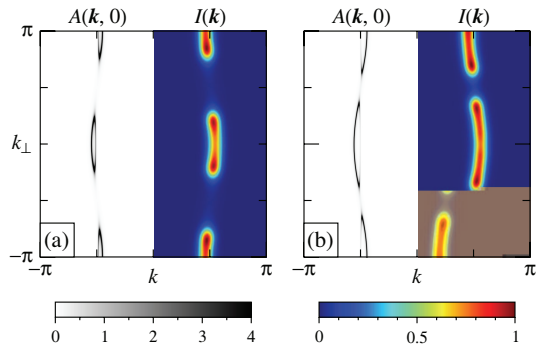}
\caption{\label{fig:ARPES}
Comparison of the zero-energy spectral function $A(\bm{k},0)$ at $T=0$ with the
expected ARPES intensity $I(\bm{k})$ calculated assuming a $k$-space resolution
of $0.04\pi$, an energy resolution $0.004\,t$, and an energy integration window
$\delta E=0.01\,t$. The self-energy parameters are taken from Fig.~
\ref{fig:hot_spots}b for $t_{\perp}=0.42$ (a) and $t_{\perp}=0.46$ (b).
}
\end{figure}

We now turn to the question of the experimental observation of FS pockets. There
are several limitations which could make the observation of such pockets by
ARPES challenging, such as the finite energy and momentum resolutions, the
finite temperature at which experiments are performed, as well as the necessity
to integrate the ARPES intensity on some energy window in order to improve the
signal to noise ratio. Ideally, ARPES would measure the occupied spectrum
$A(\bm{k},\omega)f(\omega)$. In practice, however, due to the above limitations,
the measured intensity at the Fermi energy would be
$I(\bm{k})\propto\int_{-\delta E}^{\infty}d\omega\int d\varepsilon
d\bm{q}\,A(\bm{q},\varepsilon)f(\varepsilon)g(\bm{k}-\bm{q},\omega-\varepsilon)$
where $\delta E$ defines the energy integration window and $g$ is some function
describing the instrument resolution. We have calculated $I(k,k_{\perp})$ using
a Gaussian for $g$. The comparison depicted in Fig.~\ref{fig:ARPES} of the $T=0$
Fermi surface with the expected ARPES intensity clearly shows that the closing
segments of the pockets near $k=\pi/2$ would very likely be hidden in the ARPES
signal. The broad aspect of $I(\bm{k})$ as compared to $A(\bm{k},0)$ is not a
consequence of finite temperature, but of (i) the finite $k$-space resolution
combined with the fact that the pockets are very thin and (ii) the large
difference in quasiparticle weight on the two sides of the pockets, which is
obvious in $A(\bm{k},0)$ and consistent with the residues shown in
Fig.~\ref{fig:hot_spots}. Similar FS anisotropies were recently found in
cluster-DMFT studies of the 2d Hubbard model
\cite{stanescu_pockets_cluster_dmft}, suggesting that such effects are generic
to systems close to a Mott transition, and could possibly explain the ARPES
observation of FS arcs in high-$T_c$ cuprate superconductors.

\acknowledgments
We are grateful to G. Kotliar for useful discussions and to F. C. Zhang and P.
W. Phillips for letting us know about their work \cite{yang_2006,
stanescu_preprint} as this paper was being written. This work was supported by
the Swiss National Science Foundation through Division II and the NCCR program
MaNEP, as well as by CNRS and Ecole Polytechnique.


\begin{thebibliography}{21}
\bibitem{lambda}
The parameter $\lambda$ is not needed in principle since the exchange
renormalization is already taken into account up to second order in
$\Sigma_{\protect\text{int}}$. We nevertheless kept $\lambda$ as a free
parameter for the fit, which leads to a slight improvement, even though the
fitted value of $\lambda$ is always close to unity (Fig.\protect \nobreakspace 
{}\ref{fig:hot_spots}b).
\end{thebibliography}
\end{document}